\documentclass[aps,pre,reprint,longbibliography,superscriptaddress]{revtex4-1}

\usepackage{amsmath}
\usepackage{amssymb}
\usepackage{graphicx}

\usepackage{hyperref}
\usepackage{bm}
\usepackage{color}

\begin{document}


\title{Low-frequency whistler waves excited by relativistic laser pulses}

\affiliation{Beijing National Laboratory for Condensed Matter Physics, Institute of Physics, CAS, Beijing 100190, China}
\affiliation{Department of Physics and Beijing Key Laboratory of Opto-electronic Functional Materials and Micro–nano Devices, Renmin University of China, Beijing 100872, China}
\affiliation{College of Physical Science and Technology, Sichuan University, Chengdu 610065, China}
\affiliation{School of Physical Sciences, University of Chinese Academy of Sciences, Beijing 100049, China}
\affiliation{Collaborative Innovation Center of IFSA, Shanghai Jiao Tong University, Shanghai 200240, China}
\affiliation{Songshan Lake Materials Laboratory, Dongguan, Guangdong 523808, China}
\affiliation{Key Laboratory for Laser Plasmas, Ministry of Education, Shanghai Jiao Tong University, Shanghai 200240, China}

\author{Huai-Hang Song}
\affiliation{Beijing National Laboratory for Condensed Matter Physics, Institute of Physics, CAS, Beijing 100190, China}
\affiliation{School of Physical Sciences, University of Chinese Academy of Sciences, Beijing 100049, China}
\author{Wei-Min Wang}\email{weiminwang1@ruc.edu.cn}
\affiliation{Department of Physics and Beijing Key Laboratory of Opto-electronic Functional Materials and Micro–nano Devices, Renmin University of China, Beijing 100872, China}
\affiliation{Beijing National Laboratory for Condensed Matter Physics, Institute of Physics, CAS, Beijing 100190, China}
\affiliation{Collaborative Innovation Center of IFSA, Shanghai Jiao Tong University, Shanghai 200240, China}
\author{Jia-Qi Wang}
\affiliation{College of Physical Science and Technology, Sichuan University, Chengdu 610065, China}
\author{Yu-Tong Li}
\email{ytli@iphy.ac.cn}
\affiliation{Beijing National Laboratory for Condensed Matter Physics,
Institute of Physics, CAS, Beijing 100190, China}
\affiliation{School of Physical Sciences, University of Chinese Academy of Sciences, Beijing 100049, China}
\affiliation{Collaborative Innovation Center of IFSA, Shanghai Jiao Tong University, Shanghai 200240, China}
\affiliation{Songshan Lake Materials Laboratory, Dongguan, Guangdong 523808, China}
\author{Jie Zhang}
\affiliation{Beijing National Laboratory for Condensed Matter Physics, Institute of Physics, CAS, Beijing 100190, China}
\affiliation{Collaborative Innovation Center of IFSA, Shanghai Jiao Tong University, Shanghai 200240, China}
\affiliation{Key Laboratory for Laser Plasmas, Ministry of Education, Shanghai Jiao Tong University, Shanghai 200240, China}

\date{\today}

\begin{abstract}

It is shown by multi-dimensional particle-in-cell simulations that intense secondary whistler
waves with special vortex-like field topology can be excited by a relativistic laser
pulse in the highly magnetized, near-critical density plasma. Such whistler waves with lower frequencies obliquely propagate on both
sides of the laser propagation axis. The energy conversion rate from laser to whistler waves can exceed 15$\%$. Their dispersion relations
and field polarization properties can be well
explained by the linear cold-plasma model. The present work
presents a new excitation mechanism of whistler modes extending to
the relativistic regime and could also be applied in magnetically
assisted fast ignition.

\end{abstract}


\maketitle


\section{Introduction}

Whistler waves \cite{helliwell1965whistlers} were first discovered in the
ionosphere \cite{Barkhausen1919pz} with audio frequencies. Since then, as a branch of electromagnetic modes in the magnetized plasma, the whistler wave and its
excitation have been extensively studied in space plasmas \cite{Tsurutani1974jgr}, laboratory plasmas
\cite{Stenzel1999jgr,Compernolle2016ppcf}, and solid-state plasmas \cite{Legendy1964pr, Maxfield1969ajp}. Hot electrons with the sufficient temperature anisotropy are considered to be the
dominant energetic source to stimulate whistler waves in the radiation belts \cite{Fu2014jgr,Omura2011jgr}. Modulated electron
beams are the useful tools to trigger whistler waves via various wave-particle resonances \cite{Krafft1994prl,An2019pop}. Magnetic
antennas can generate whistler wave packets with helical phase fronts \cite{Stenzel2019pop}. The \v{C}erenkov whistler emission is
also found in the magnetic reconnection \cite{Goldman2014prl,Steinvall2019prl}.

Recently, the whistler wave has also aroused interest in laser-plasma interactions, particularly in fast ignition, with the emergence of unprecedented strong magnetic fields up to the kilo-tesla level \cite{Fujioka2013sr}.  The strong magnetic fields lasting a few nanoseconds were experimentally produced at the center of coil targets driven by high-power nanosecond laser pulses \cite{Fujioka2013sr,Santos2018pop,Zhang2018hplse}. With the assistance of such an external magnetic field, high-energy electron beams can be well guided in a high-density fusion fuel and the heating efficiency to the fuel can be significantly enhanced in the fast ignition, as shown in both simulations \cite{Strozzi2012pop,Wang2015prl} and experiments \cite{Bailly2018nc,Sakata2018nc}.  Higher laser energies and optimized coil targets are expected to further increase the  external magnetic field strength. When the external magnetic field exceeds a critical value $B_c=m_ec\omega_0/|e|$ (in Gaussian units), which means that the non-relativistic electron cyclotron frequency $\omega_{ce}$ exceeds the laser frequency $\omega_0$, the laser-plasma interactions enter the whistler-mode regime, where $m_e$ is the electron rest mass and $e$ is the electron charge. In this case, the laser pulse can penetrate into an overdense plasma for a long distance due to the absence of cutoff density \cite{Yang2015apl,Luan2016pre}. Based on this unique property, a few theoretic works have shown the enhanced electron heating \cite{Wu2017ppcf,Gong2017pop}, cyclotron resonance absorption \cite{Sano2017pre}, and wave to ion energy transfer \cite{Sano2019pre,sano2020pre} in this regime. In these works, direct whistler-mode conversion was investigated in the highly magnetized overdense plasma that the laser pulse was considered as a whistler wave with the unchanged frequency before and after entering the plasma. In addition to this kind of direct whistler-mode conversion, intense laser interactions with the magnetized plasma may excite diverse and complicated secondary whistler waves of broad frequencies, for example, through whistler instabilities induced by electron flows \cite{Gary2005book,Taguchi2017jpp}.

In this paper, we investigate low-frequency, secondary whistler wave excitation in the interaction of a relativistically intense laser pulse with a highly magnetized near-critical density plasma. For the laser penetration in overdense plasmas \cite{Wu2017ppcf,Gong2017pop,Sano2017pre,Sano2019pre,sano2020pre}, this kind of secondary whistler wave emissions should also be taken into account since a lower-density preplasma usually is unavoidable. Our two-dimensional (2D) and three-dimensional (3D) particle-in-cell (PIC) simulations show that the excited whistler waves have special vortex-like field structures. The dispersion relation and field polarization properties of excited whistler waves are examined using a linear cold-plasma model.

In Sec.~\ref{simulation}, we present the PIC simulation results to show the field topology and time evolution of the excited whistler waves. In Sec.~\ref{analysis}, we give a qualitative picture about the whistler excitation mechanism. In addition, the dispersion relation as well as field polarization of whistler waves is analyzed based on a cold-plasma model. In Sec.~\ref{parameters}, we discuss the impacts of parameters of the laser pulse, plasma density, and external magnetic field on the whistler wave excitation, and a 3D simulation is carried out. Section~\ref{conclusion} contains a brief conclusion.

\section{Simulation results}
\label{simulation}

To study the whistler wave excitation, we carry out a series of 2D and 3D relativistic PIC simulations. The main results are given by 2D PIC simulations due to a lower computational expense. Then, a 3D simulation is performed to further confirm the field topology of excited whistler waves.

\begin{figure}[t]
\centering
\includegraphics[width=3.4in]{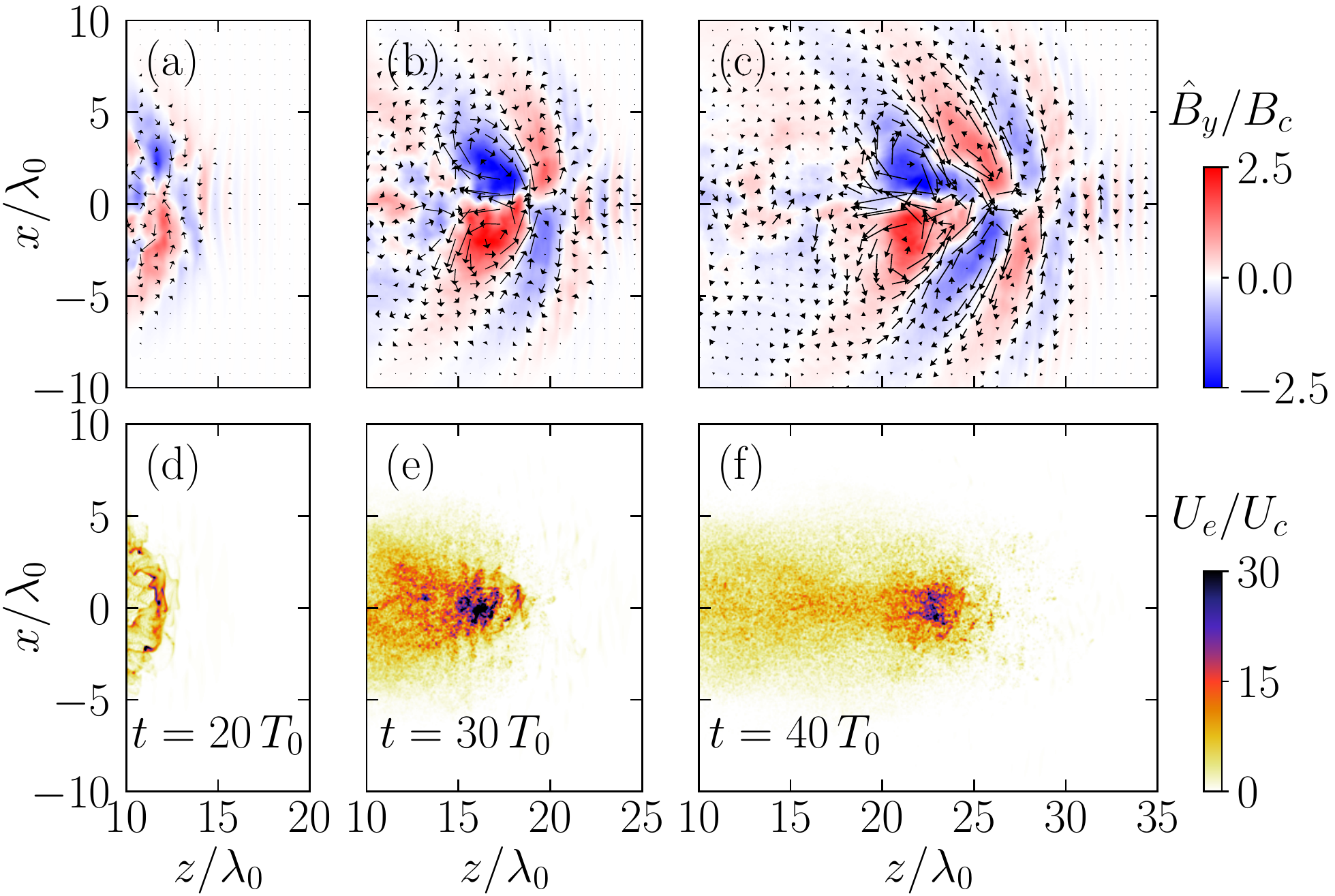}
\caption{\label{fig1} Snapshots of azimuthal magnetic field $\hat B_y$ (top row) and energy density of electrons $U_e$ (bottom row) at three different times [(a), (d)] $t=20T_0$, [(b), (e)] $t=30T_0$, and [(c), (f)] $t=40T_0$, where $U_c$ is $n_cm_ec^2$. Vector plots of the magnetic field ($\hat B_x$, $\hat B_z$) in the $x$-$z$ plane are also presented in (a)-(c) to show the field topology.}
\end{figure}

In the first 2D simulation, a laser pulse with a profile of $a_L=a_0\,{\rm sin}^2(\pi t/\tau_0)\times{\rm exp}{(-r^2/r^2_0)}$ within $0< t \leq\tau_0$ is normally incident from the left boundary (at $z=0$). We take the laser normalized peak strength $a_0=|e|E_L/m_ec\omega_0=5.0$, spot size $r_0=4\lambda_0$, and pulse duration $\tau_0=20T_0$, where $T_0=2\pi/\omega_0$ is the laser period, $\lambda_0=2\pi/k_0$ is the laser wavelength, and $k_0$ is the laser wavenumber. At the initial moment $t=0$, the plasma is composed of cold electrons and protons ($m_p/m_e=1836$).  The computational domain has a size of $60\lambda_0\times 110\lambda_0$ in $x\times z$ directions with $1920\times 3520$ cells. A vacuum is located at $0<z\leq5\lambda_0$, a low density plasma with an exponential ramp from $0.01 n_c$ to $0.8 n_c$ is within $5<z\leq10\lambda_0$, and a uniform plasma with a density of $n_0=0.8n_c$ is followed at $z>10\lambda_0$, where $n_c=m_e\omega_0^2/4\pi e^2$ is the critical density. A uniform external magnetic field with a magnitude $B_0=3.0B_c$ is imposed along the laser pulse propagation direction (i.e., along the $+z$ axis). Each cell contains 16 macroparticles for each species. To reduce the numerical heating, the fourth-order interpolation \cite{Esirkepov2001cpc} is applied. Absorbing boundary conditions are used for both particles and fields in any direction.



\begin{figure}[t]
\centering
 \includegraphics[width=3.0in]{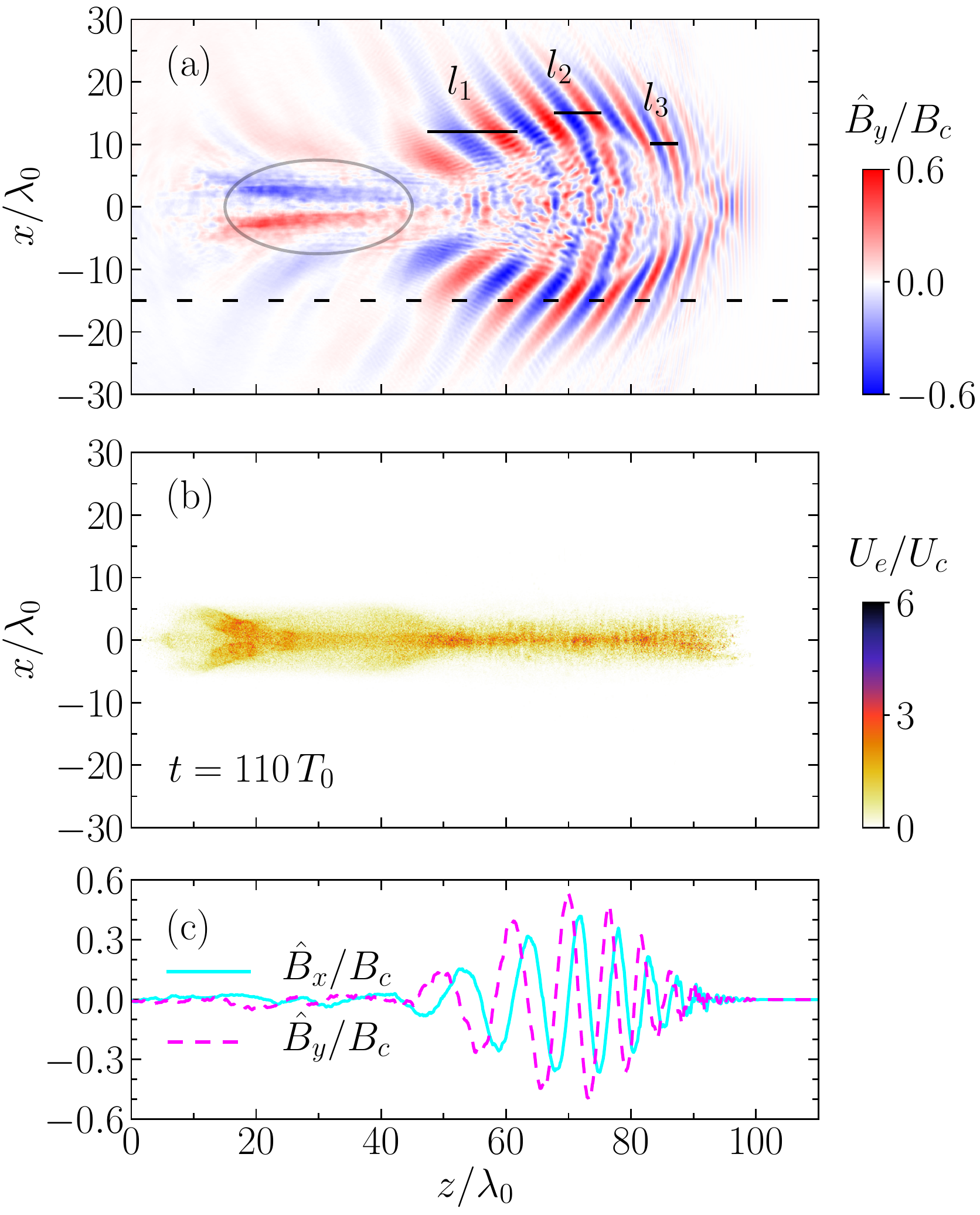}
\caption{\label{fig2} Snapshots of (a) azimuthal magnetic field $\hat B_y$ and (b) electron energy density $U_e$ at $t=110T_0$. We choose three local whistler segments along the $z$ direction, as indicated by the solid lines $l_1$, $l_2$, and $l_3$ in (a), to further analyze the field polarization in detail in Fig.~\ref{fig6}. (c) One-dimensional spatial profiles of $\hat B_x$ and $\hat B_y$ components taken from $x=-15\lambda_0$ as indicated by the dashed line in (a).}
\end{figure}

Figures~\ref{fig1}(a)-\ref{fig1}(c) show the spatial distributions of the out-of-plane or azimuthal magnetic field $\hat B_y$ and in-plane magnetic field vector ($\hat B_x, \hat B_z$) at different times, displaying the formation of whistler waves as the laser pulse penetrates into the plasma. Here, we add a hat symbol $\wedge$ over a variable to mean that the relevant quantity is averaged over one laser period in order to filter out the fast-varying laser field. Note that the axial magnetic component $\hat B_z$ or $B_z$ excludes the external magnetic field $B_0$. It can be seen that strong whistler waves are excited with a strength of $2.5B_c$ in the $\hat B_y$ component, which approaches the external magnetic field strength of $3.0B_c$. The whistler waves are right-hand circularly polarized through the field linkage between $\hat B_y$ and ($\hat B_x, \hat B_z$) as shown in Figs.~\ref{fig1}(b) and \ref{fig1}(c), similar to spheromak-like field perturbations observed in the laboratory plasma \cite{Stenzel2006prl,Eliasson2007prl}. In addition, the excited whistler waves have a cone-shaped phase structure due to their oblique propagation along the external magnetic field. The complete 3D field topology will be shown by a 3D PIC simulation later in Fig.~\ref{fig10}. In Figs.~\ref{fig1}(d)-\ref{fig1}(f), we display the corresponding energy density distribution of electrons. Electrons in the laser interaction zone are rapidly accelerated up to tens of MeV. These high-energy electrons are strongly confined in the transverse direction by the external magnetic field.

\begin{figure}[t]
\centering
\includegraphics[width=3.1in]{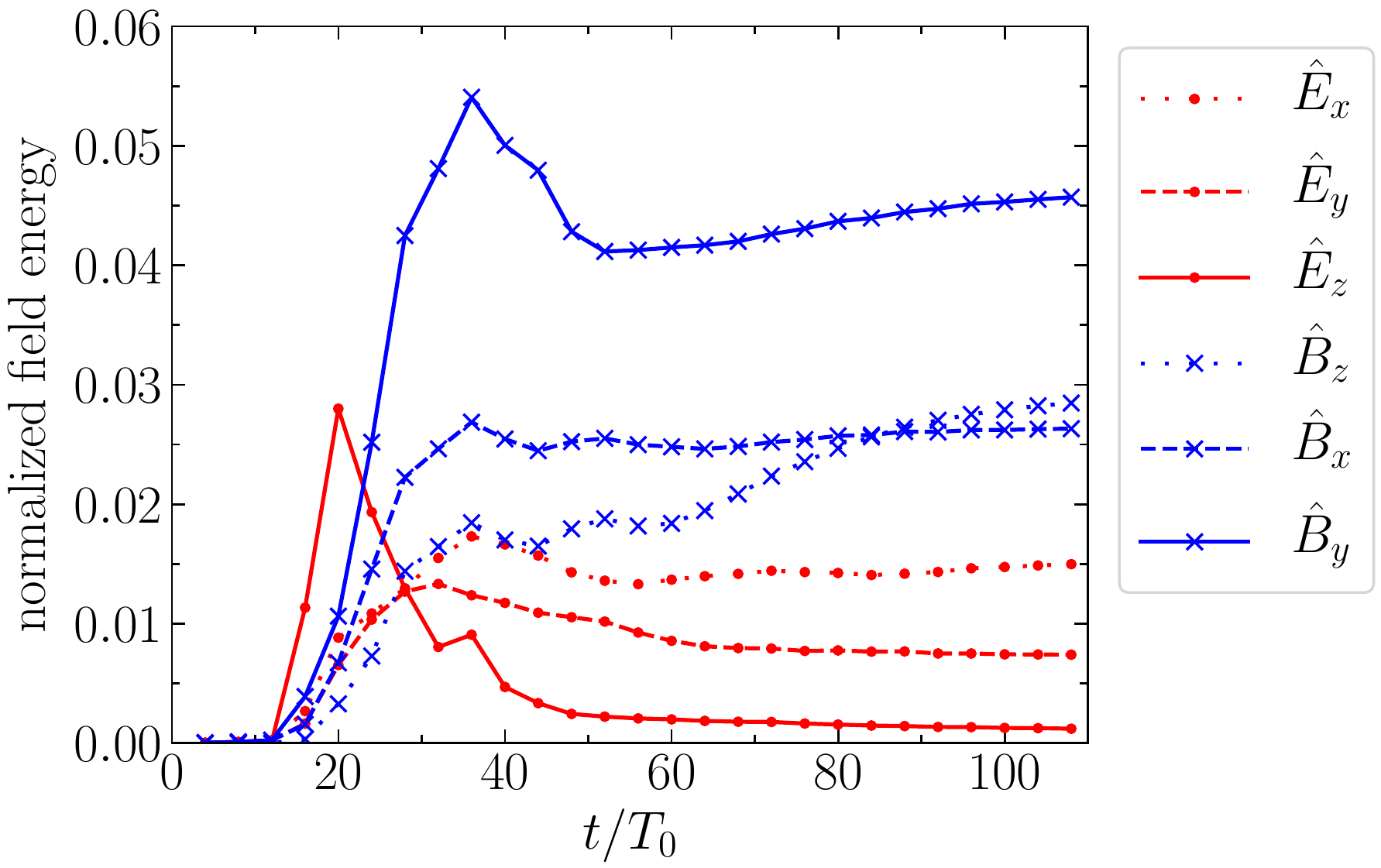}
\caption{\label{fig3}Temporal evolution of energy of the electric and magnetic fields averaged by one laser period. Each energy component is normalized by the total laser energy.}
\end{figure}



The whistler waves gradually develop to have more wave cycles with time and higher-frequency components go ahead, as shown in Fig.~\ref{fig2}(a) at a later time of $t=110T_0$. This is  because the higher-frequency whistler waves have higher group velocities in the plasma. Stable whistler waves obliquely propagate on both sides of the laser propagation axis, with wide local wave normal angles in the range of $-50^\circ$ and $50^\circ$. The local wave normal angle is defined as the one between the local whistler wave vector $\bm{k}$ and external magnetic field direction. We show a one-dimensional magnetic field distribution along $x=-15\lambda_0$ in Fig.~\ref{fig2}(c). The handedness of $\pi/4$ phase shifting between $\hat B_x$ and $\hat B_y$ confirms its right-hand polarization with respect to the external magnetic field. Figure~\ref{fig2}(b) is the corresponding electron energy density, showing that hot electrons are mainly confined around the laser propagation axis within a narrow zone by the strong external magnetic field. Obliquely propagating whistler waves spread in a broader zone where electrons have low energies and are still cold. One can analyze the whistler dynamics by a cold plasma model, as we will do in Sec.~\ref{analysis}. At the center area, the whistler field pattern looks turbulent and complex due to high-energy electron dynamics.

Figure~\ref{fig3} gives the energy evolution of one-laser-period-averaged (low-frequency) field components with time. The laser pulse impinges the plasma in the interval $10T_0<t<30T_0$, leading to a significant increase of whistler wave energy. Meanwhile, the whistler wave with its low frequency and high intensity can be depleted by the hot plasma since it is basically near the laser interaction zone in this time. As whistler wave generation and depletion are roughly counteracted, its field energy reaches the maximum at about $t=30T_0$. After $t=50T_0$, the residual energy of each field component remains nearly constant with time, suggesting the stable, non-dissipative whistler wave propagation in the cold plasma region. The whistler wave carries about $9\%$ of the total laser energy. The energy of magnetic fields ($\sim7\%$) is relatively higher than that of electric fields ($\sim2\%$), where the energy of axial electric component $\hat E_z$ is almost zero due to a very high conductivity along the external magnetic field direction. The slight increases of $\hat B_y$ and $\hat B_z$ energies after $t=60T_0$ are attributed to quasi-static magnetic field growth in the left high-energy electron region as marked by an elliptic curve in Fig.~\ref{fig2}(a), independent of propagating whistler waves.

\begin{figure}[t]
\centering
 \includegraphics[width=3.2in]{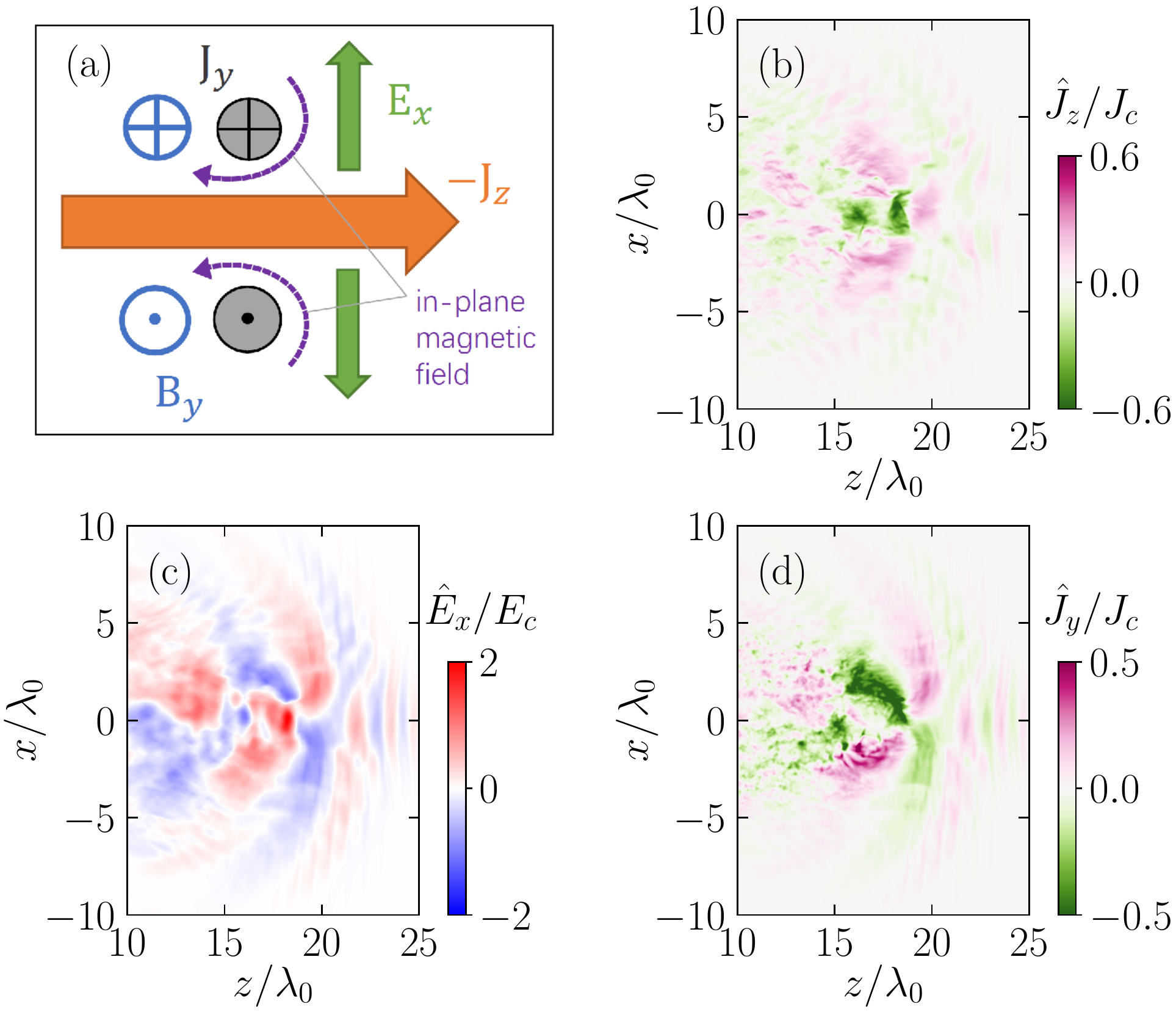}
\caption{\label{fig4}(a) Schematic of whistler wave excitation. Snapshots of (b) axial current $\hat J_z$, (c) radial electric field $\hat E_x$, and (d) azimuthal current $\hat J_y$ at $t=30T_0$ from the simulation, where $E_c$ is $m_ec\omega_0/|e|$ and $J_c$ is $n_c |e|c$. The azimuthal magnetic field component $\hat B_y$ is induced by the strong axial current $\hat J_z$, and the in-plane magnetic field component ($\hat B_x, \hat B_z$) is induced by the azimuthal Hall current $\hat J_y$.}
\end{figure}

\section{Theoretical analysis}
\label{analysis}

The whistler wave excitation can be understood in
electron magnetohydrodynamic regime by a simple 2D physical
picture as sketched in Fig.~\ref{fig4}(a). When a relativistically intense laser pulse
propagates through a near-critical plasma, a ion channel can be formed
\cite{Pukhov1996prl,Pukhov1999pop} by the laser pondermotive force. This pondermotive force
pushes electrons both forward and laterally, resulting in a strong
axial current $\hat J_z$, azimuthal magnetic field $\hat B_y$, and radial
electric field $\hat E_x$, as shown in Figs.~\ref{fig4}(b), \ref{fig1}(b), and \ref{fig4}(c), respectively. With an external magnetic field
along the laser propagation direction, an azimuthal Hall current
$\hat J_y$ shown in Fig.~\ref{fig4}(d) is also driven by the radial electric field $\hat E_x$ through
$\bm{E}\times \bm{B}_0$ electron drifts, hence inducing the in-plane
magnetic field as shown in Fig.~\ref{fig1}(b). The superposition of azimuthal
and in-plane magnetic fields forms the observed vortex-like field structure. The
similar whistler excitation by the electrode \cite{Urrutia1994pop} or magnetic antenna
\cite{Stenzel1993pf} in non-relativistic regime has also been observed in discharge plasmas.

Then, a linear theory based on a cold magnetized plasma approximation is adopted here to study the properties of excited whistler waves, due to the fact that the whistler waves are obliquely propagating and mainly located outside the laser interaction zone where electrons have low energies. In the model, we ignore the motion of ions and collision effects, which is appropriate in the interaction between a relativistically ultrashort laser pulse and near-critical plasma. Because the whistler waves discussed here have broad spectra and their high-frequency components are close to the plasma frequency $\omega_{pe}=\sqrt{4\pi n_e e^2/m_e}$, the widely used dispersion formula of whistler modes $c^2k^2=\omega\omega_{pe}^2/(\omega_{ce}\cos\theta-\omega)$ cannot hold well, particularly in the present case of $\omega_{pe}\lesssim\omega_{ce}$ \cite{helliwell1965whistlers}. Here, $\theta$ is the local wave normal angle and $\omega_{ce}=|e|B_0/m_ec$. In our case with the external magnetic field along the $+z$ axis and the whistler wave propagates in the $x$-$z$ plane, the general wave equation can be given by \cite{Stix1992book}

\begin{equation}\label{eq1}
\left(
\begin{array}{ccc}
S-n^2\cos^2\theta & -iD & n^2\cos\theta\sin\theta
\\iD & S-n^2 & 0
\\n^2\cos\theta\sin\theta & 0 & P-n^2\sin^2\theta
\end{array}
\right)
\left(\begin{array}{c}E_x\\ E_y\\E_z\end{array}\right)=0
\end{equation}
where $S=\frac{1}{2}(R+L)$, $D=\frac{1}{2}(R-L)$, $P=1-\omega_{pe}^2/\omega^2$, $R=1-\omega_{pe}^2/(\omega^2-\omega \omega_{ce})$, $L=1-\omega_{pe}^2/(\omega^2+\omega \omega_{ce})$, and $n=ck/\omega$ is the refractive index. Note that $\omega_{ce}$ is a positive value in this paper.

A nontrivial solution of Eq.~\ref{eq1} requires that the determinant of $3\times 3$ matrix is zero, which yields the well-known dispersion relation \cite{Stix1992book}
\begin{equation}\label{eq2}
n^2=\frac{B\pm F}{2A},
\end{equation}
where $A=S\sin^2\theta+P\cos^2\theta$, $B=RL\sin^2\theta+PS(1+\cos^2\theta)$, and $F^2=(RL-PS)^2\sin^4\theta+4P^2D^2\cos^2\theta$.

\begin{figure}[t]
\centering
\includegraphics[width=2.6in]{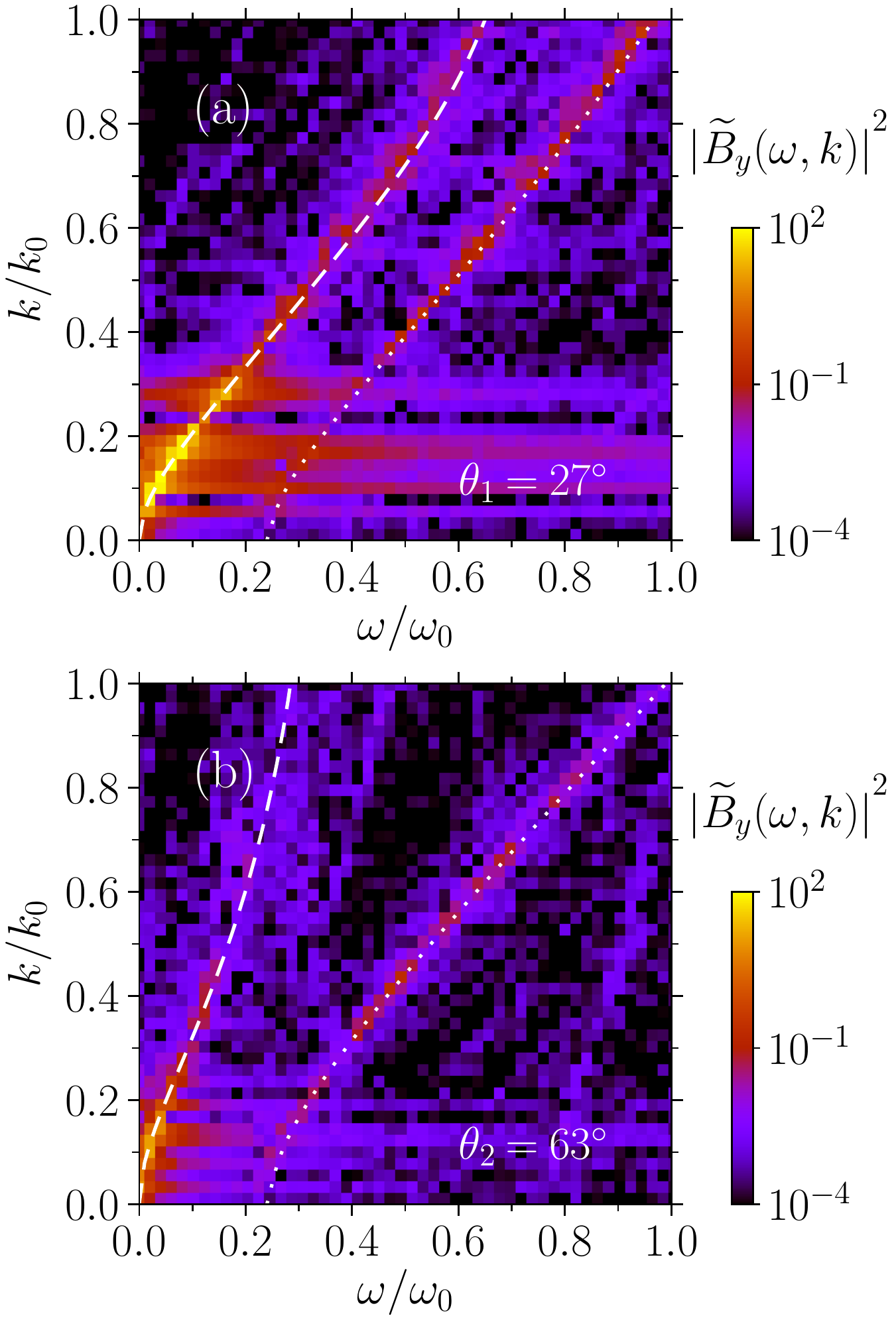}
\caption{\label{fig5} The power spectral density of $B_y$ at two selected wave normal angles (a) $\theta_1=27^\circ$ and (b) $\theta_2=63^\circ$ in the $\omega$-$k$ coordinates, obtained from our PIC simulation. In each plot, the dashed curve stands for the theoretical whistler-mode dispersion relation given by Eq.~\ref{eq2} with $\omega_{pe}=\sqrt{0.8}\omega_0$ and $\omega_{ce}=3.0\omega_0$ at the corresponding angles. The dotted curve stands for the theoretical dispersion relation of the left-hand wave.}
\end{figure}

To gain insight into the dispersion properties of excited whistler modes in our PIC simulation, we perform 3D fast Fourier transformation of magnetic field $B_y$ in the region of $10\lambda_0<z<100\lambda_0$ over the time domain of $70T_0<t<110T_0$ to obtain the power spectral density $|\widetilde B_y(\omega,\bm{k})|^2$ in the $\omega-\bm{k}$ space. Here $k_x$ and $k_z$ are the wave vector components along $x$ and $z$ directions, respectively, and they satisfy $k^2=k_x^2+k_z^2$ and ${\rm tan}\,\theta=k_x/k_z$. We select two $\omega-k$ spectral slices of wave
normal angles $27^\circ$ and $63^\circ$ from the
$\omega-\bm{k}$ space, as shown in Figs.~\ref{fig5}(a) and ~\ref{fig5}(b). The whistler modes at these two angles agree well with the theoretical dispersion relation (dashed curves) governed by Eq.~\ref{eq2}. The simulation further shows that this magnetized plasma system also supports left-hand electromagnetic modes with frequencies higher than $\frac{1}{2}\left(\sqrt{\omega_{ce}^2+4\omega_{pe}^2}-3\omega_0\right)=0.246\omega_0$. These much weaker left-hand modes also match the theoretical dispersion relation (dotted curves).

\begin{figure}[t]
\centering
\includegraphics[width=2.9in]{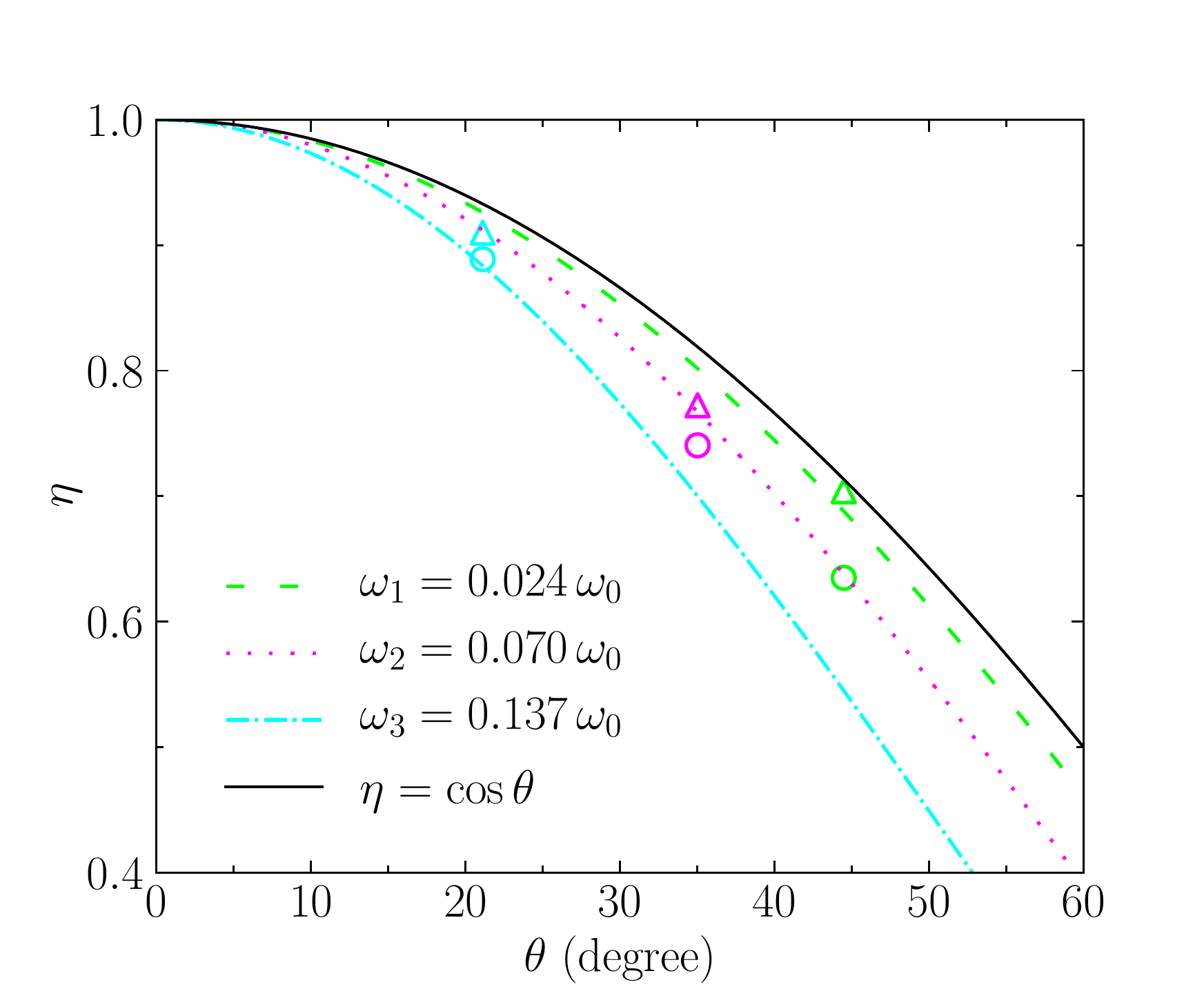}
\caption{\label{fig6} The theoretical $\eta$ as a function of $\theta$ according to Eq.~\ref{eq1} at three different wave frequencies $\omega_1=0.026\omega_0$ (dashed green curve), $\omega_2=0.072\omega_0$ (dotted violet curve), $\omega_3=0.156\omega_0$ (dot-dashed cyan curve), where $\omega_{pe}=\sqrt{0.8}\omega_0$, $\omega_{ce}=3.0\omega_0$. $\omega_{1}$, $\omega_{2}$ and $\omega_{3}$ correspond to frequencies of three whistler segments $l_1$, $l_2$ and $l_3$ indicated in Fig.~\ref{fig2}(a), respectively. The limit solution $\eta=\cos\theta$ is also plotted as a reference (black solid curve). Triangle and circle marks represent $|B_x|/|B_y|$ and $|E_y|/|E_x|$ values of these three whistler segments from the simulation with $l_1$ in green, $l_2$ in violet, and $l_3$ in cyan.}
\end{figure}

The wave polarization information in terms of electric field $\bm E$ can be given by Eq.~\ref{eq1}. Through simple operations, the ratios of electric field components can be characterized by
\begin{align}\label{eq3}
    \frac{E_y}{E_x} &=i\frac{D}{n^2-S},
    &
    \frac{E_z}{E_x} &=\frac{n^2\sin\theta\cos\theta}{n^2\sin^2\theta-P}.
\end{align}
With Faraday's law $\triangledown \cdot {\bf E}=-\frac{1}{c}\partial {\bf B}/\partial t$, the ratios of the magnetic field components can be derived as
\begin{align}\label{eq4}
\frac{B_z}{B_x} &=-\tan\theta,
&
\frac{B_y}{B_x} &=\tan\theta\frac{E_z}{E_y}-\frac{E_x}{E_y}.
\end{align}
The ratio of $B_z$ to $B_x$ is completely determined by Maxwell's equations, thus it is a universal relation independent of the plasma behavior. Equation~\ref{eq3} and \ref{eq4} can be further simplified with $-P\gg 1$ and one can obtain $E_z/E_x\approx0$ and $B_y/B_x\approx-E_x/E_y$. It implies that longitudinal electric field $E_z$ is negligible, which is consistent with the result in Fig.~\ref{fig3}. With these simplifications and defining the dimensionless parameter $\eta=\left|D/(n^2-S)\right|$, the amplitude ratios of field components can be approximated as follows:

\begin{align}\label{eq5}
\frac{|B_x|}{|B_y|}&\approx \frac{|E_y|}{|E_x|}=\eta,
&
\frac{|E_z|}{|E_x|}&\approx0,
&
\frac{|B_z|}{|B_x|}&=|\tan\theta|.
\end{align}
The limits of small wave frequencies $\omega\ll \omega_{pe}$ and $\omega_{ce}<\omega_{pe}$ are well satisfied in typical magnetospheric plasmas and the relation $\eta\approx|\cos\theta|$ is obtained \cite{Verkhoglyadova2010jgr,Bellan2013pop}.
In this situation, the magnetic field is still circularly polarized even for obliquely propagating whistler waves, while the electric field is circularly polarized only in the direction transverse to the wave vector. In our case with $\omega_{ce}>\omega_{pe}$, the relation $\eta\approx|\cos\theta|$ still holds but with a higher requirement for the limit of $\omega\ll \omega_{pe}$, as shown in Fig.~\ref{fig6}.

\begin{figure}[t]
\centering
\includegraphics[width=2.7in]{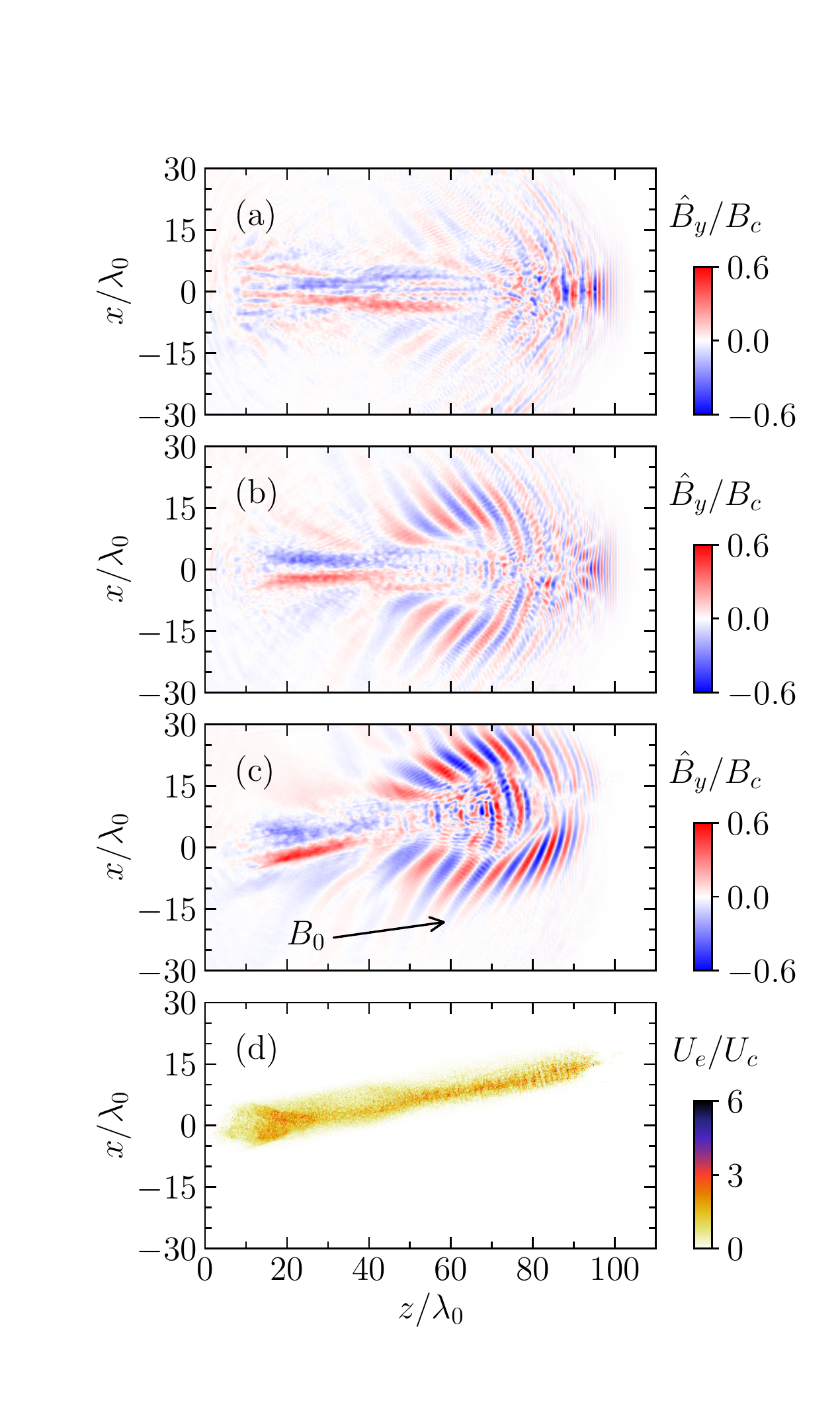}
\caption{\label{fig7} Snapshots of azimuthal magnetic field $\hat B_y$ at $t=110T_0$, where the laser and external magnetic field parameters are identical to those of Fig.~\ref{fig2}, but for (a) a left-hand circularly polarized laser, (b) a linear polarized laser, and (c) the obliquely imposed external magnetic field with an angle of $10^\circ$ to the $+z$ axis. (d) Energy density of electrons $U_e$ corresponding to (c).}
\end{figure}

To examine the whistler wave polarization, we select three single-cycle whistler segments at different positions as indicated by lines $l_1$, $l_2$, and $l_3$ in Fig.~\ref{fig2}(a). Each line is parallel to the $x$ axis. For each whistler segment, we treat it as a plane wave, so Eq.~\ref{eq5} can be applied. The local wave normal angle can be inferred through the ratio of $|B_z|$ to $|B_x|$ in Eq.~\ref{eq5} due to its independence on specific electromagnetic modes. In practice, we use the sum of absolute values of each field segment to calculate their ratios. We estimate the wavenumber component $k_z$ directly from Fig.~\ref{fig2}(a) and then the wavenumber $k=k_z/{\rm cos}\,\theta$ can be obtained. The refractive index $n$ and frequency $\omega$ can be calculated from the dispersion relation Eq.~\ref{eq2}, since it has been confirmed by Fig.~\ref{fig5}. The theoretic curves of $\eta$ corresponding to each whistler segment are shown in Fig.~\ref{fig6}. The field component ratios $|B_x|/|B_y|$ and $|E_y|/|E_x|$ obtained from the PIC simulation are marked by triangles and circles, respectively, in a good agreement with the theoretic curves. The ratio of $|E_y|/|E_x|$ from each field segment is lower than the theoretical value, mainly because the $|E_x|$ component is affected by electrostatic field under above estimates.

\begin{figure*}[t]
\centering
\includegraphics[width=6.3in]{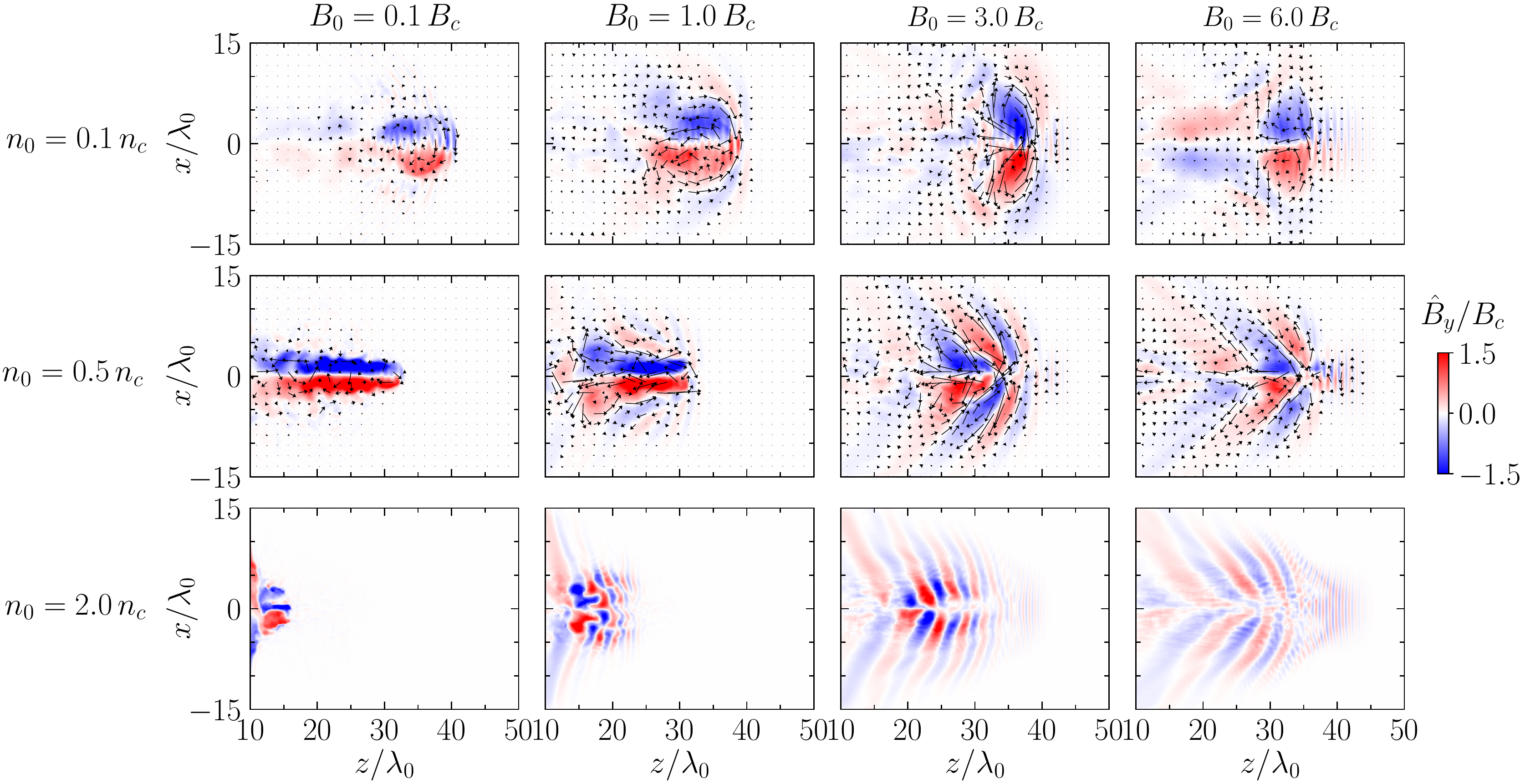}
\caption{\label{fig8} Snapshots of azimuthal magnetic field $\hat B_y$ at $t=50T_0$ under different external magnetic field strengths $B_0=0.1B_c$ (first column), $1.0B_c$ (second column), $3.0B_c$ (third column), and $6.0B_c$ (fourth column), as well as different plasma densities $n_0=0.1n_c$ (first row), $0.5n_c$ (second row), and $2.0n_c$ (third row). Vector plots of the magnetic field ($\hat B_x$, $\hat B_z$) in the $x$-$z$ plane are also presented in the first two rows.}
\end{figure*}

\section{impacts of laser, plasma, and magnetic-field parameters}
\label{parameters}

We proceed to investigate the impact of laser parameters on the whistler wave excitation. First, the laser pulse should be relativistically intense ($a_0>1$). Below the relativistic strength, the laser field behavior follows the linear dispersion relation given by Eq.~\ref{eq2}, suggesting that secondary emissions with frequencies different from the laser frequency cannot be triggered.


The laser polarization also plays a crucial role. Above, we have taken the right-hand circularly polarized laser pulse, and consequently strong right-hand  circularly polarized whistler waves are observed. When we change the laser polarization to be left-hand in Fig. \ref{fig7}(a), the low-frequency field looks much weaker compared with that of Fig. \ref{fig2}(a). The high dependence on laser polarization may seem strange because whistler waves are driven by the electron currents due to laser pondermotive force as described in Fig. \ref{fig4}. One possible reason is that the right-hand circularly polarized laser pulse interacts more strongly with the magnetized plasma than the left-hand laser, hence it can drive more strong electron currents to excite whistler waves. The complex interaction among laser field, plasma, and whistler wave in the relativistic region is still to be studied. We further set the laser pulse to be linearly polarized along the $x$ axis with an amplitude of $a_0=5\sqrt2$ to keep the total laser intensity unchanged. In Fig. \ref{fig7}(b), the excited whistler waves are stronger than those by the left-hand polarized laser pulse in Fig. \ref{fig7}(a), but weaker than those by the right-hand one in Fig. \ref{fig2}(a), since a linear polarization can be decomposed into a left-hand polarization and a right-hand one.

In Figs. \ref{fig7}(c) and \ref{fig7}(d), we take the external
magnetic field has a angle  $10^\circ$ with respect to the laser
propagation direction (the $+z$ axis). The excited whistler waves
have a similar pattern to the co-directed configuration shown in
Fig. \ref{fig2}(a), but the pattern axis is along the direction of the
external magnetic field, rather than the laser propagation
direction. This can be explained by the generated hot electrons
strongly confined along the external magnetic field direction, as
shown in Fig. \ref{fig7}(d). Therefore, the central axis of the whistler wave
pattern is mainly determined by the external magnetic field.

\begin{figure}[b]
\centering
\includegraphics[width=3.0in]{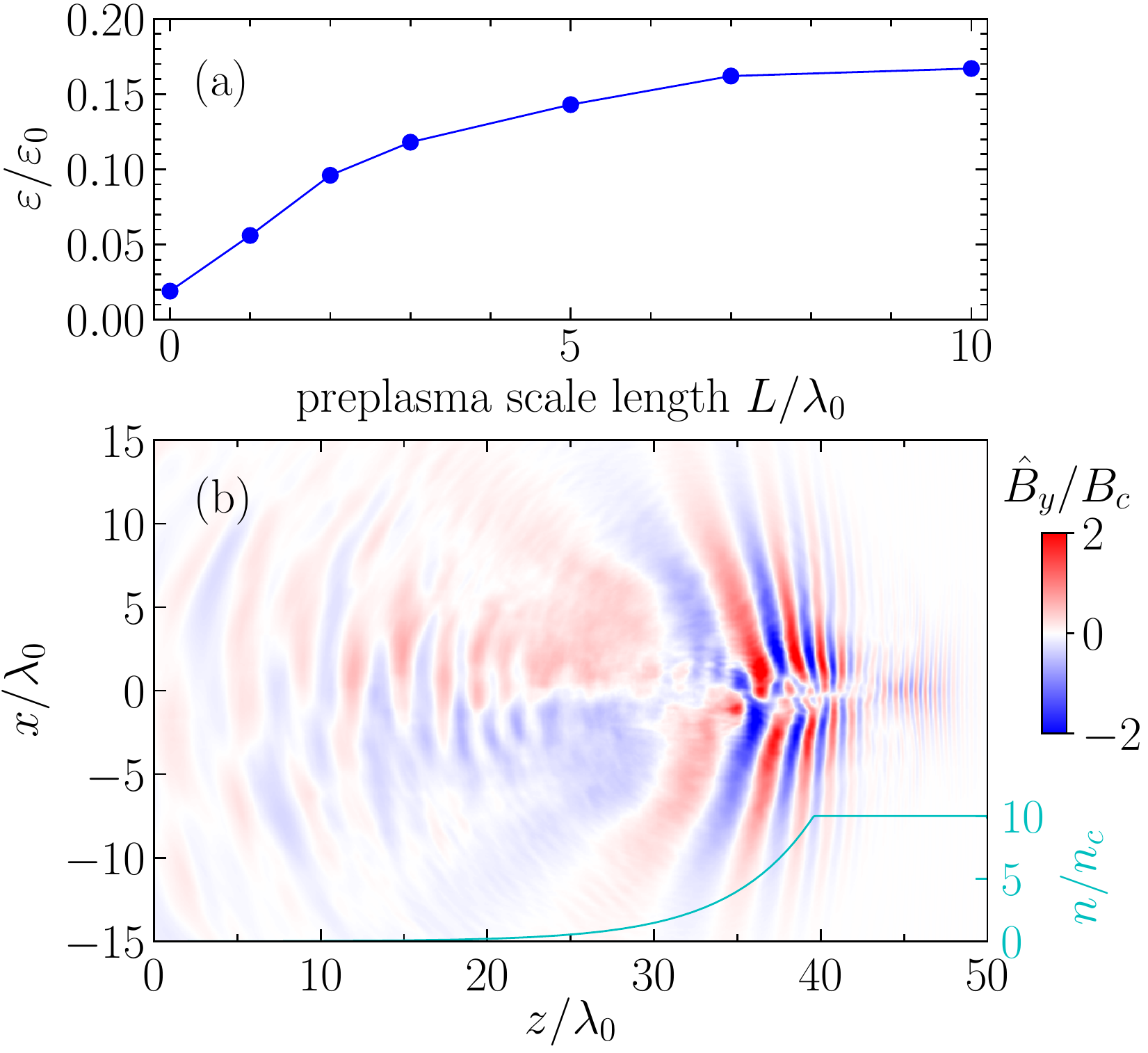}
\caption{\label{fig9}(a) The low-frequency whistler conversion rate $\varepsilon/\varepsilon_0$ versus preplasma scale
lengths $L$, where $\varepsilon$ and $\varepsilon_0$ are the
whistler wave energy and the total laser energy, respectively. (b) Snapshot of azimuthal magnetic field $\hat B_y$ at $t=50T_0$ in the case of $L=5\lambda_0$. The spatial profile of initial plasma density $n$ is also given.}
\end{figure}


\begin{figure*}[t]
\centering
\includegraphics[width=4.8in]{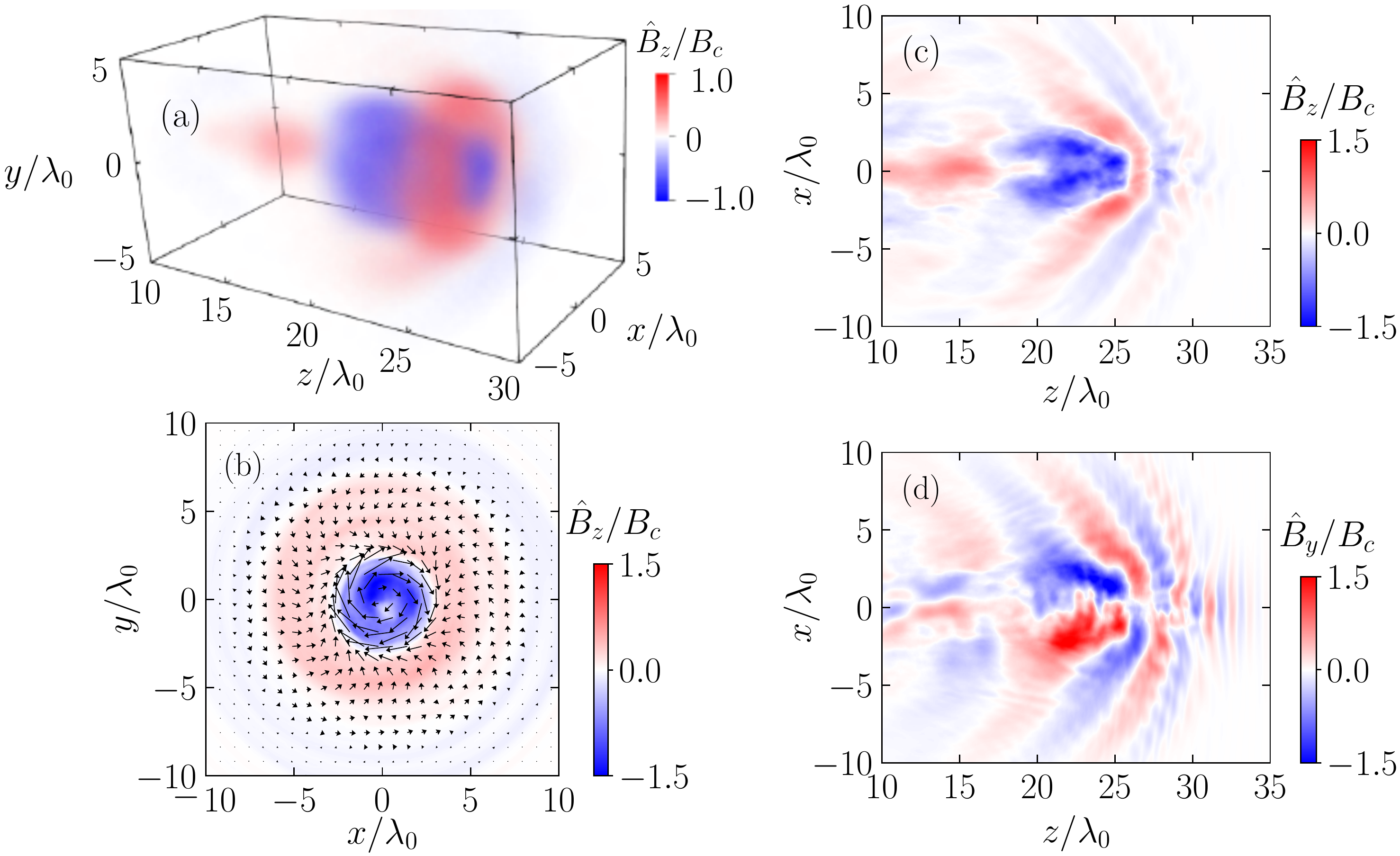}
\caption{\label{fig10} 3D simulation results at $t=30~T_0$.
(a) Profile of axial magnetic field $\hat B_z$ in the 3D volume view. (b) Snapshot of axial magnetic field $\hat B_z$ associated with the vector plot of ($\hat B_x$, $\hat B_y$) at $z=22\lambda_0$ in the $x$-$y$ plane. [(c), (d)] Snapshots of $\hat B_z$ and $\hat B_y$ at $y=0$ in the $x$-$z$ plane.}
\end{figure*}
In Fig.~\ref{fig8}, we scan the plasma density $n_0$ and external magnetic field strength $B_0$, while keeping other parameters the same as those in Fig.~\ref{fig2}. For a plasma density much lower than the critical one (see $n_0=0.1n_c$ cases in Fig.~\ref{fig8}), a stable spheromak-like whistler wave packet can be formed, which is different from the typical multi-cycle whistler waves in a higher density plasma at the same time (see $n_0=0.5n_c$ or $n_0=2.0n_c$ cases in Fig.~\ref{fig8}), but similar to their early states (see the $n_0=0.8n_c$ case in Fig.~\ref{fig2}(b)). Besides, the excited
whistler waves are weaker as the plasma density increases to over
$n_c$ (see $n_0=2.0n_c$ cases in Fig.~\ref{fig8}). This illustrates
advantages of near-critical density for whistler wave excitation and
agrees with the below results in Fig.~\ref{fig9} that the whistler wave
conversion is much lower in the absence of low-density preplasma. By comparing these cases under different external magnetic field strengths, we can find that strong whistler waves can be excited only when $B_0$ exceeds $B_c$. Below the critical strength $B_c$, the generated azimuthal magnetic fields are primarily quasi-static fields, not waves. Note that the condition $B_0>B_c$ is also required for the laser penetration in overdense plasma \cite{Wu2017ppcf,Gong2017pop,Sano2017pre,Sano2019pre,sano2020pre}. From this point of view, this kind of secondary whistler wave emission should also be taken into account even in the study of overdense plasmas if a relativistic intense laser is employed, since a low-density preplasma at the front surface usually is unavoidable, particularly in fast ignition \cite{Li2008prl}.

To illustrate the influence of preplasma on the low-frequency whistler wave
conversion in the interaction of laser pulse with strongly
magnetized overdense plasmas, we take an overdense plasma with a
density of $10n_c$ and in the front of it there is a preplasma with
an exponential density profile of a scale length $L$. Other
simulation parameters are the same as Fig. \ref{fig2}. The energy of excited
low-frequency whistler waves normalized by the total laser energy is
shown in Fig. \ref{fig9}(a). For a steep overdense interface of $L=0$, the
conversion ratio is less than 0.02. Taking into account relativistic effects, the corrected critical density in this case is increased to
$n_c^*=n_c\sqrt{1+a_0^2}\approx 5n_c$, hence there is still a small
whistler conversion ratio when $L=0$ and $n_0=10n_c \approx 2n_c^*$.
The conversion ratio is expected to be much less than 0.02 for a
higher-density plasma without the preplasma. With increasing the preplasma
scale length, the conversion ratio rises up dramatically, and it saturates up to 0.16 at $L=10\lambda_0$. Thus, the excitation process of low-frequency whistler waves in the preplasma region should be taken into account, at least in terms of energy share. Figure \ref{fig9}(a) shows the azimuthal magnetic field component $\hat B_y$ in a representative case of $L=5\lambda_0$, we can see strong whistler waves are excited in the preplasma and then they propagate deep into the overdense plasma region due to the absence of cutoff density.

Finally, we conduct a 3D PIC simulation, to demonstrate the 3D field topology of whistler waves. We take a simulation box of $30\lambda_0\times 30\lambda_0\times 30\lambda_0$ in $x\times y\times z$ directions with $480\times 480\times 480$ cells. Each cell contains 8 macroparticles for each species. Other parameters are still the same as those in Fig.~\ref{fig2}. The field distribution of axial magnetic component $\hat B_z$ at $t=30T_0$ via a 3D volume view is shown in Fig.~\ref{fig10}(a). Its two central slices in the $x$-$y$ plane and $x$-$z$ plane are also shown in Fig. \ref{fig10}(b) and \ref{fig10}(c), respectively. It can be seen that the axial component $\hat B_z$ is almost axisymmetric with respect to the $z$ axis. In Fig.~\ref{fig10}(b), we also plot the azimuthal field vector ($\hat B_x$, $\hat B_y$). The azimuthal magnetic field has a right-hand linkage with respect to the axial magnetic field. A central slice of $\hat B_y$ in Fig.~\ref{fig10}(d) is almost identical to that in Fig.~\ref{fig1}(c) that we have obtained in the 2D PIC simulation.

To excite strong low-frequency whistler waves, the required external magnetic field strength should be higher than the critical magnetic field $B_c$ from our simulations, which is determined by the laser wavelength $\lambda_0$. For Ti:Sapphire lasers with a typical wavelength $\lambda_0=800$ nm, $B_c\approx 13~{\rm kT}$, which is about one order of magnitude higher than that achieved in present experiments \cite{Fujioka2013sr,Santos2018pop,Zhang2018hplse}.
For CO$_2$ lasers \cite{Haberberger2010oe} with a wavelength of $10~\mu{\rm m}$, the critical magnetic field can be significantly reduced to a more realistic value of $B_c\approx 1~{\rm kT}$.

\section{conclusion}
\label{conclusion}

In summary, a series of 2D and 3D PIC simulations show
that a relativistic laser pulse can excite low-frequency, vortex-like
whistler waves in a highly magnetized, near-critical density plasma. The excited whistler modes at different
wave normal angles obtained from PIC simulations are in
agreement with  the dispersion relation given by a cold-plasma
theory. The theory also confirmed the field polarization properties. By scanning parameters
of plasma density and external magnetic field strength, we find that the
whistler waves can be widely excited if $B_0>B_c$, which is the
same as the requirement for the direct whistler-mode conversion of
laser pulses in overdense plasmas. This work enriches the whistler
wave excitation mechanism and extends it to the relativistic
interaction regime. These results could be referred in the
magnetically assisted fast ignition study and a double-cone ignition
(DCI) project recently funded in China \cite{Zhang2020rsta}.


\begin{acknowledgments}This work was supported by the Strategic Priority Research Program of Chinese Academy of Sciences (Grant
Nos. XDA25050300, XDA25010300, and XDB16010200), the National Key R\&D Program of China (Grant No. 2018YFA0404801), National
Natural Science Foundation of China (Grant Nos. 11775302, 11827807 and 11520101003), Science Challenge Project of China (Grant No.
TZ2016005), the Fundamental Research Funds for the Central Universities, the Research Funds of Renmin University of China (20XNLG01), Sichuan Science and Technology Program No.2017JY0224, and the Innovation Spark Project of Sichuan University No. 2018SCUH0090.
\end{acknowledgments}

\bibliography{whistler_waves}

\end{document}